\documentstyle[preprint,aps]{revtex}
\begin{document}
\draft
\title{\bf Salty Water Cerenkov Detectors for Solar Neutrinos}
\author{W.C. Haxton}
\address{Institute for Nuclear Theory, Box 351550 \\
and Department of Physics, Box 351560 \\
University of Washington, Seattle, Washington  98195-1550}
\date{\today}
\maketitle
\begin{abstract}
The addition of certain solutes to a water Cerenkov detector will introduce
new charge-current channels for the detection of $\nu_e$s.  The experimental
conditions necessary to exploit such signals - large volumes and very
low backgrounds - should be reached for the first time in SuperKamiokande and
SNO.  I compare some of the more attractive possibilities ($^7$Li,
$^{11}$B, $^{35}$Cl) in terms
of counting rate, reliability of the cross section, specificity of signal,
and potential for distinguishing competing solutions to the
solar neutrino puzzle.
\end{abstract}
\pagebreak
The purpose of this letter is to stimulate discussions of additional uses of
water Cerenkov detectors in solar neutrino physics.  The introduction of
relatively inexpensive solutes can produce new charge-current signals that can
be combined with $\nu_x - e$ elastic scattering to determine the spectra of
both electron and heavy-flavor neutrinos.  The advances being made by
SuperKamiokande [1] and the Sudbury Neutrino Observatory (SNO) [2] -
extraordinary fiducial
volumes and radiopurities - are essential to such schemes.  The discussion
below of three
possible solutes will illustrate what can be achieved in terms of counting
rates, specificity of the signal, and reliability of cross section - as well
as the kinds of compromises that experimenters must consider when
attempting to exploit such signals in solar neutrino experiments.

Some of the advantages of deuterium, the neutrino target in SNO, are also
offered by the series of light ``n - p" systems, those stable mirror
nuclei with T = 1/2.  Quite often the cross sections can be determined
very accurately.
For light targets (e.g., $^7$Li and $^{11}$B), the thresholds are quite
low, so that rates, especially
per target nucleon, are very large.  Alternatively, in the heaviest cases
(e.g., $^{31}$P, $^{35}$Cl, and $^{39}$K), the Coulomb interaction has
produced thresholds for $(\nu_e, e^-)$ that are sufficiently high that the
subsequent $\beta^+$ decay of the daughter nucleus might be observable,
providing a distinctive coincidence in time and position.

In terms of counting rate and reliability of cross section calculations, the
outstanding example among the light mirror systems is $^7$Li$(\nu_e,
e^-)^7$Be.
Only two transitions are
significant, the mixed Fermi (F) - Gamow Teller (GT) analog transition to
the $^7$Be ground
state $(E_\nu^{thresh} = 0.861$ MeV) and the GT transition of superallowed
strength to the 0.478 MeV first excited state $(E_\nu^{thresh}$ = 1.339 MeV).
The allowed matrix elements (Table 1) can be extracted from the electron
capture rates for $^7$Be.  Thus the cross section for a normalized $^8$B
neutrino spectrum is known to about
1\%, $\sigma(^8$B) = 3.50 $\cdot 10^{-42}$ cm$^2$.  Because of the low
threshold and the
concentration of the transition strength near threshold,
properties $^7$Li shares with deuterium, the
electron spectrum is particularly sensitive to distortions in the
incident neutrino spectrum.  I will return to this issue later.

A number of lithium-bearing solutes could be considered, several of which
(e.g., LiCl) are very soluble.  For definiteness I will consider a 5\%
solution of LiOH, a modestly soluble compound (12.8 g/100 cc at 20$^\circ$C)
containing no other principal isotopes that interact with solar $\nu_e$s.  For
the 22
kiloton fiducial volume of SuperKamiokande and an undistorted $^8$B $\nu_e$
flux consistent with the results [3] of Kamioka II/III (3 $\cdot$
10$^6$/cm$^2$
sec, or about half the standard solar model [4] (SSM) result), the total
event rate is
8480/year, comparable to the $\nu_x$-electron inelastic scattering rate.  For
a threshold trigger on the apparent electron energy of $E_A >$ 5 MeV and a
Gaussian resolution function [1] of $\sigma$(E) = 0.16E $\sqrt{{10 MeV
\over E}}$,
87.4\% of these events would be detected.

In the early planning stages of SUNLAB [5] a concentrated ($\agt$ 30\%)
solution
of LiCl was investigated but this option was abandoned because of the
associated severe light attenuation, especially in the UV.  A naive
extrapolation of the measurements suggests this
problem would not arise for the much lower
concentrations considered here.  That is, the large volume of SuperKamiokande
makes a Li solute feasible.  The absence of an attenuation problem at low
concentrations must be established experimentally, of course.

There is one unfortunate aspect of $^7$Li: the F and GT angular distributions,
proportional to $1 + \hat{\nu} \cdot \vec{\beta}_e$ and $1 - {1 \over 3}
\hat{\nu} \cdot \vec{\beta}_e$,
respectively, conspire to produce a nearly isotropic cross section for $^8$B
neutrinos
\begin{equation}
\sigma \,\, \propto (1 + \alpha(E_e) \hat{\nu} \cdot \vec{\beta}_e)
\eqnum{1}
\end{equation}
where $\hat{\nu}$ is the unit vector in the direction of the neutrino,
$\vec{\beta}_e = \vec{p}_e/E_e$, and $\alpha(E_e) \sim -0.02 \pm 0.02$ for
5 MeV
$\alt E_e \alt$ 10 MeV.  (The precise variation with $E_e$ can be computed
from the results in Table 1.)  Thus, unlike $\nu_x - e$ elastic scattering,
the angular distribution cannot be used to separate neutrino events from an
isotropic background.  This constrains useful measurements
to those larger electron energies where background rates are
negligible.

For this reason, one might consider other solutes where the angular
distribution is more distinctive.  Among the light mirror nuclei
$^{11}B(\nu_e,e^-)^{11}$C, with a threshold of 1.982 MeV, is probably the best
alternative to $^7$Li.  Boron was discussed previously as a solar
neutrino target in the proposal Borex [6], a predecessor to Borexino.

The nuclear physics is summarized in Table 1.  The F-GT transition to the
$^{11}$C ground state carries 73\% of the $^8$B $\nu_e$ absorption strength.
Thus the reaction is again reasonably hard.  The transition strength is fixed
by the $^{11}$C lifetime.  The first excited state accounts for 16\% of the
cross section.  Its strength is also tightly constrained by experiment, as
the $\gamma$-decay rate from this state to the ground state is known
in both $^{11}$B and $^{11}$C.  This allows one to extract
\begin{equation}
| \langle J_f \| \, \sum^A_{i =1} \, (\mu_1 \vec \sigma (i) + \vec \ell (i))
\,\, \tau_3 \,\,
(i) \| \, J_i \rangle | \eqnum{2}
\end{equation}
where $\mu_1$ = 4.706.  The large isovector magnetic moment means Eq. (2)
is dominated
by the GT contribution, which can be extracted by doing a shell model
calculation of the matrix element ratio $\eta = \langle \ell (i) \tau_3 (i)
\rangle/\langle \mu_1 \vec \sigma (i) \tau_3 (i)\rangle$.  Since $\eta$ is
small
(-0.144 in a 1p-shell calculation [7]), the nuclear structure uncertainties in
this extraction are quite modest.  The result (Table 1) is consistent with
direct shell model estimates of the GT matrix element, corresponding to an
effective $F_A^{eff}$ = 0.94 near the expected value.

Essentially all of the remaining 10\% of the cross section is carried by the
5/2$^-$(4.32 MeV) and 3/2$^-$(4.80 MeV) second and third excited states in
$^{11}$C. The GT strengths in Table 1 are shell model [7] results:
calculations
analogous to that described above cannot be made because the $\gamma$-decay
lifetimes in $^{11}$C have not been measured.  If they were, and if we assign
a pessimistic error to shell model estimates of $\eta$ of $\pm$ 50\%,
the uncertainty in the total cross section
$\sigma(^8$B) = 1.41 $\cdot 10^{-42}$ cm$^2$ could be reduced to less than
$\pm$ 5\%.
Thus $^{11}$B is another case where nuclear structure uncertainties
are under reasonably good control.

A possible solute is the weak acid H$_3$BO$_3$, with a solubility of 6.35
g/100
cc at 20$^\circ$C.  Assuming a 5\% solution, the 22 kiloton fiducial volume of
SuperKamiokande, and an undistorted $^8$B $\nu_e$ flux of 3 $\cdot$
10$^6$/cm$^2$ sec, the total event rate is 1430/year.  For the SuperKamiokande
threshold trigger $E_A$ = 5 MeV and resolution function, 60\%
of these events would be detected.

The angular distribution is again precisely calculable using the results of
Table 1.  An approximation to this result, in the notation of Eq. (1), is
\begin{equation}
\alpha(E_e) = 0.31 + 0.063 \, (E_e - 7.5) \eqnum{3}
\end{equation}
where $E_e$ is in MeV.  The cross section is modestly forward peaked, with
$\sigma (0^\circ)/\sigma(180^\circ) \sim 1.9$ at $E_e$ = 7.5 MeV.
As in the case of $\nu_x - e$ scattering,
this angular dependence can be exploited to distinguish
an important fraction of neutrino events from the isotopic background.
However this more specific signal comes at a price:  the event rate,
relative to
$^7$Li, is significantly lower, and the higher threshold (by 1.1 MeV) somewhat
restricts the range of neutrino energies that can be sampled.

A very different experimental strategy is possible with heavy mirror
systems.  Because the introduction of $^{40}$K in a water Cerenkov
detector would be inadvisable, the best case may be
\begin{equation}
\nu_e + ^{35} \mathrm{Cl} \rightarrow e^- + ^{35} \mathrm{Ar} \eqnum{4a}
\end{equation}
\begin{equation}
^{35} \mathrm{Ar} \rightarrow ^{35} \mathrm{Cl} + e^+ + \nu_e \, , \,
\tau_{1/2} = 1.775 ~\mathrm{sec}
\eqnum{4b}
\end{equation}
where the maximum positron energy is 5.454 MeV.  The large threshold
produces a potentially distinctive neutrino reaction signal:  a prompt
electron followed by a positron correlated in time ($\sim$1.775 sec delay) and
position.  The time distribution must follow the $^{35}$Ar decay curve.
Furthermore, there is very significant peaking of the prompt electrons
in the forward direction because the F strength dominates the
analog transition: $\alpha(E_e) \sim 0.86$ in Eq. (1).  All of these
considerations conspire to make the signal unusually clean.

The cross section can be determined precisely from the $^{35}$Ar
$\beta$ decay lifetime and measured decay branches.  The superallowed ground
state transition accounts for 94\% of the transition strength.  The remaining
6\%, according to standard 2sld shell model calculations [8], is carried by
transitions calibrated by $^{35}$Ar $\beta$ decay.  Detailed results are given
in Table 1.  The substantial threshold leads to a cross section
$\sigma(^8B) = 2.23
\cdot 10^{-43}$ cm$^2$, considerably smaller than those in $^{11}$B and $^7$Li.

While a number of highly soluble Cl salts might be used, NH$_4$Cl
is an attractive choice because it can be synthesized from very high purity
gases [9].  A 10\% solution would produce 400 events/year in a
22 kiloton volume, assuming a $^8$B neutrino flux of $3 \cdot 10^6$/cm$^2$ sec.
Event detection is governed by the prompt trigger threshold
(5 MeV) and the sensitivity to the delayed $\beta^+$.  The current
Kamioka III data acquisition system is capable down to 1.6 MeV,
with triggering efficiencies becoming sufficiently high ($\agt$ 95\%) by
2.5 MeV
that the spectrum above this energy is undistorted [3,10].  Thus this is a
reasonable threshold for the positron trigger.  Folding these
thresholds with the anticipated SuperKamiokande resolution yields
efficiencies of 43.7\% and 62.2\% for the prompt electron and coincident
positron, respectively.  Thus the number of detected events from
such a concentration of Cl in SuperKamiokande would be
109/year, comparable to the $\nu_x - e$ rate of Kamiokande III.

The key issue, whether such a modest signal will be swamped by
accidental coincidences, was recently explored by
members of the SuperKamiokande collaboration [10].  For the expected
vertex resolution ($\pm$ 1m for low-energy events) and background
rates (assuming the anticipated 250-fold decrease [10] in the radon
content relative to Kamiokande III), the signal/accidental rate
is $\sim$ 1.  This is very satisfactory, given the angular and
time distributions that will distinguish real events from background.
As the rate of accidental coincidences depends quadratically on the
radon concentration, it is clear that the substantially improved
radiopurity of SuperKamiokande is essential to any such neutrino
detection scheme.

The addition of one of the discussed solutes to
SuperKamiokande will allow experimeters to compare electron and heavy-flavor
neutrino signals as a function of energy.  This is a capability not
presently envisioned for
either SuperKamiokande or SNO: the SNO neutral current signal,
neutrons from the neutrino breakup of deuterium, measures an integrated
rate.  Neutral current sensitivity is provided by the $\nu_x$ - e reaction:
the electron and heavy-flavor cross sections are in the
approximate ratio 7:1.  The comparison of the charge-current reaction
off the solute to $\nu_x$ - e scattering would not be influenced by
uncertainties in the electron detection efficiency.

To illustrate what might be learned, I have investigated four scenarios
that would produce identical $\nu_x$ - e rates under the operating
conditions of
Kamioka II/III: $i$) an undistorted $^8$B flux reduced to about 50\% of
the SSM result ($3 \cdot 10^6$/cm$^2$sec); $ii$) the ``small angle"
Mikheyev-Smirnov-Wolfenstein (MSW) solution [11] for flavor oscillations
(sin$^2 2 \theta = 0.005, \delta m^2 = 3.46 \cdot 10^{-6}$ eV $^2$);
$iii$) the ``large-angle"
flavor oscillation solution (sin$^2 2 \theta = 0.825, \delta m^2 =
10^{-5}$ eV $^2$);
and $iv$) the small-angle sterile oscillation solution (sin$^2 2 \theta =
0.005, \delta m^2 = 2.63 \cdot 10^{-6}$ eV $^2$).  Figure 1 shows the event
distribution, in 1 MeV bins, for the solute concentrations discussed
earlier, a 3-year running time, and the envisioned experimental conditions
for SuperKamiokande.  Figure 2 shows the ratio of nuclear and $\nu_x$ - e
event rates, normalized in each bin to the results for scenario $i$).

{}From Fig. 1a one sees that the ability of SuperKamiokande to
distinguish between these scenarios using $\nu_x$ - e data depends heavily
on the results
from the lowest energy bins (below 7 MeV); that the effects
are less than 10\%, so that one must know the absolute
efficiency of the detector at the few percent level; and that
scenario $i$) cannot be distinguished from $iii$), nor $ii$)
from $iv$) with any significant degree of confidence.
The $^7$Li results are quite interesting in this connection.
Under the anticipated initial operating conditions [10] of SuperKamiokande,
the signal/background rates in this channel will be approximately
3, 20, and 150 for events collected above 10, 11, and 12 MeV, respectively.
The 3-year rates above these thresholds, 6680, 4095, and 2240
events for scenario $i$), respectively, are still quite reasonable.
This reaction nicely complements $\nu_x$ - e scattering:
the ratio of $^7$Li to elastic scattering events (Fig. 2a) differs
by more than 20\% between scenarios $i$) and $iii$), and by about
17\% between $ii$) and $iv$).  These scenarios could be distinguished
at confidence levels of approximately 8, 6, and 4$\sigma$, respectively,
using the data above 10, 11, and 12 MeV (assuming negligible backgrounds
in each case).  These errors are dominated by the statistics for
$\nu_x$ - e, which has the lower event rate in the highest energy bins.

The ratio of $^{11}$B($\nu,e^-)^{11}$C and $\nu_e$ - e events provides a
similar constraint.  The results shown in Fig. 2c were obtained by
assuming that a portion of the $^{11}$B events - those contributing
to the angular variation in the cross section - could be separated
from a larger, isotropic background in a maximum liklihood analysis.
Including all events above the 5 MeV SuperKamiokande threshold,
the statistics are then sufficient to distinguish scenario $i$) from
$iii$), or $ii$) from $iv$), at a confidence level of about 5$\sigma$.
A similar calculation for the ratio of $^{35}$Cl and $\nu_x$ - e scattering
events (Fig. 2d) yields a confidence level of about 3$\sigma$, reflecting the
limited $^{35}$Cl event rate.

These calculations, while exploratory, demonstrate that certain low-cost
solutes
could significantly enhance the power of SuperKamiokande to test
competing solutions of the solar neutrino puzzle.  The case of $^7$Li
appears especially interesting due to its precisely known cross
section and low threshold, which leads to large event rates for
electron energies above 10 MeV, where SuperKamiokande backgrounds
should be small.  Hopefully the arguments presented here
will stimulate experimenters to consider the technical issues associated
with such solutes, including light attenuation, possible chemical
effects on detector materials, solute-associated backgrounds,
and background subtractions that might be made by varying solute
concentrations or exploiting the earth's orbital variations.

I thank R. Davis, Jr., K. Innoe, L. Peak, H. Robertson, Y. Suzuki, J.
Wilkerson, and
especially Y. Totsuka for helpful communications.  This work was supported in
part by the U.S. Department of Energy.

Email address:  haxton@phys.washington.edu

\pagebreak

\begin{table}
\caption{Allowed nuclear matrix elements $|$M$|^2$ = BGT + BF and the resulting
cross sections, averaged over an undistorted $^8$B neutrino flux.}
\begin{tabular}{lllll}
Target&$E_f$(MeV)&BGT&BF&$\sigma(^8$B) (10$^{-42}$ cm$^2$) \\ \hline
$^7$Li&0.0&1.747&1.00&2.299$^{1)}$ \\
&0.478&1.630&&1.198$^{1)}$ \\
&$\geq$6.68&&&0.003$^{2)}$ \\
{}~~total&&&&3.500 \\
$^6$Li&0.0&2.580&&0.057$^{1),5)}$ \\
&$\geq$1.67&&&0.002$^{2),5)}$ \\
{}~~total&&&&0.059$^{5)}$ \\
$^{11}$B&0.0&0.601&1.00&1.025$^{1)}$ \\
&2.000&0.706&&0.229$^{3)}$  \\
&4.319&0.681&&0.079$^{2)}$ \\
&4.804&0.847&&0.076$^{2)}$ \\
&$\geq$8.1&&&0.003$^{2)}$ \\
{}~~total&&&&1.412 \\
$^{10}$B(total)&$\geq$3.35&&&0.009$^{2),6)}$ \\
$^{35}$Cl&0.0&0.114&1.00&0.209$^{1)}$ \\
&1.18-4.09&0.419$^{4)}$&&0.014$^{1),4)}$ \\
&$\geq$4.6&&&0.000$^{2)}$ \\
{}~~total&&&&0.223

\end{tabular}

$^{1)}$ Normalized to known mirror $\beta$ decay rates \\
$^{2)}$ Shell model estimates [7,8] \\
$^{3)}$ Determined from $\gamma$ decay rates (see text) \\
$^{4)}$ Sum of six allowed transitions \\
$^{5)}$ Results multiplied by $^6$Li/$^7$Li abundance ratio  \\
$^{6)}$ Results multipled by $^{10}$B/$^{11}$B abundance ratio

\end{table}

\pagebreak

\pagebreak

\begin{figure}
\caption{Number of events, in one MeV bins, for the indicated
targets and exposures, after folding with the anticipated resolution
of SuperKamiokande.  Each grouping of four bars corresponds, from
the left to right, to scenarios $i$) through $iv$) (see text).
The darkened portions indicated the $\pm 1 \sigma$ statistical errors.
In each case, the last bin shown contains all events of that energy
and above.}
\label{fig1}
\end{figure}

\begin{figure}
\caption{
The ratio of the solute and $\nu$-e event rates, shown
in one MeV bins and normalized in each bin to the results for
scenario $i$).  The labelling is as in Fig. 1.  In Fig. 2c, only
the nonisotropic component of the $^{11}$B event rate has been retained.}
\label{fig2}
\end{figure}

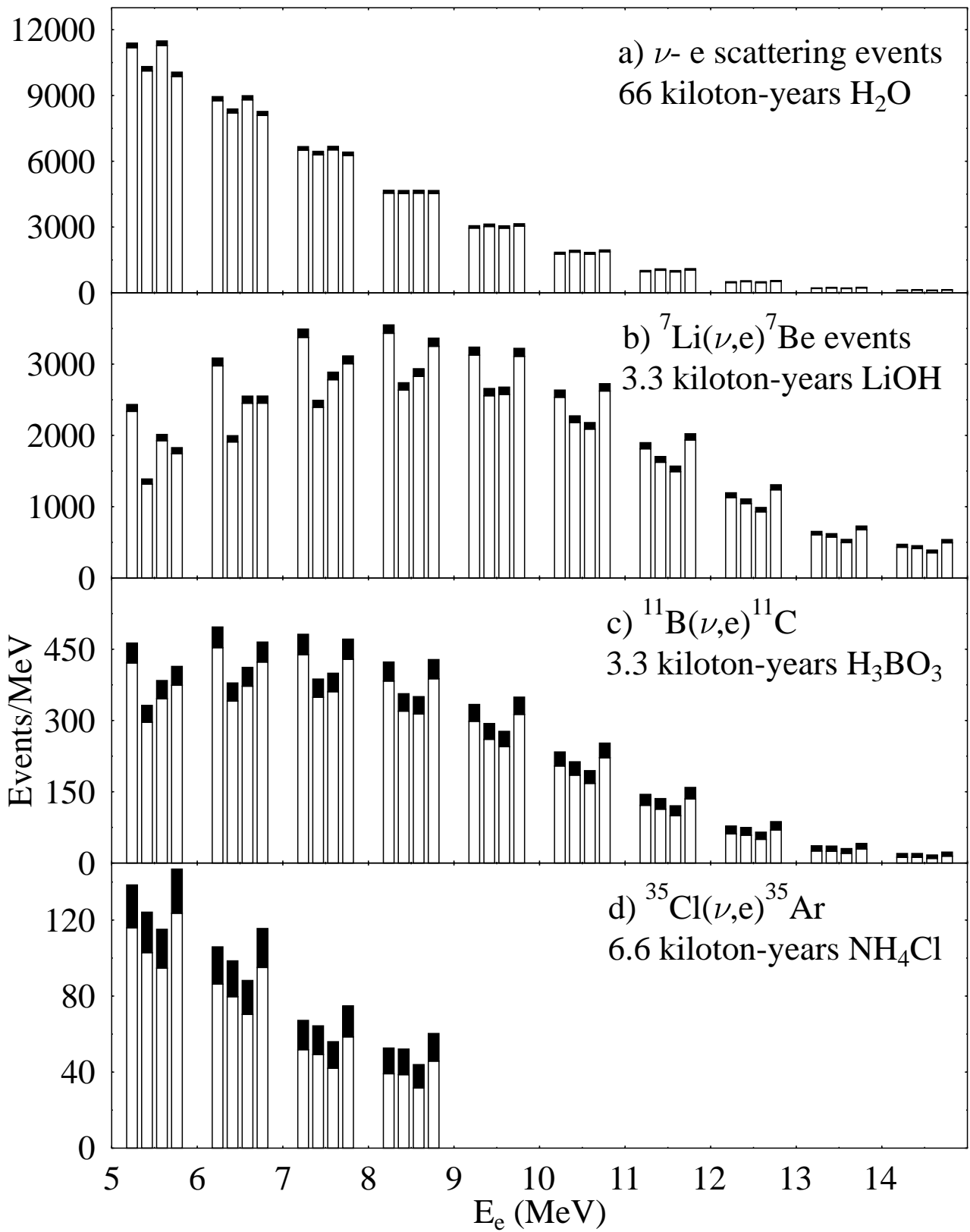
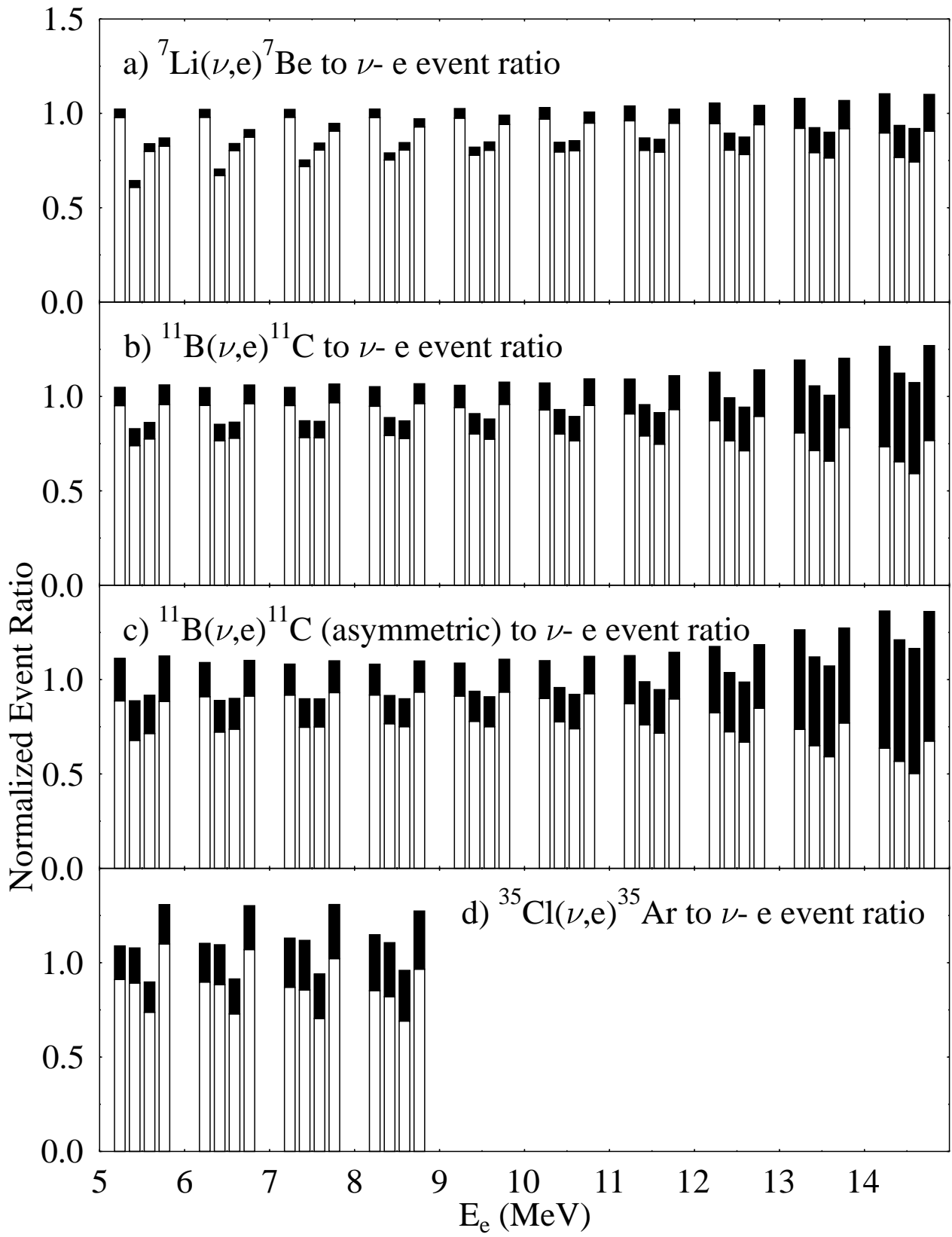
\end{document}